\documentclass[11pt,a4paper]{article}
 \pdfoutput=1
\usepackage{jcappub}
\usepackage{amsmath}
\usepackage{amsfonts,color}
\usepackage{amssymb,float}
\usepackage{mathtools}
\usepackage[utf8]{inputenc}
\usepackage{url}
\usepackage{graphicx}
\usepackage{refstyle}
\usepackage{wasysym}
\usepackage{tabularx}
\usepackage{accents}
\usepackage{graphicx}
\usepackage{color}
\usepackage[dvipsnames]{xcolor}
\usepackage{hyperref}

\hypersetup{
    colorlinks=true,
    linkcolor=blue,
    filecolor=magenta,      
    citecolor=red
}

\usepackage{subfigure}		
\usepackage{hyperref} 
\usepackage[normalem]{ulem} 
\usepackage{multirow}

\makeatletter
\long\def\dddddot#1{%
  {\mathop {#1}\limits ^{\vbox to-1.4\ex@ {\kern -\tw@ \ex@ \hbox {\normalfont .....}\vss }}}%
}
\long\def\multidots#1#2{%
  \count@=0
  {{\mathop {#2}\limits ^{\vbox to-1.4\ex@ {\kern -\tw@ \ex@ \hbox {\normalfont %
  \loop%
  \ifnum#1>\count@%
  .%
  \advance\count@ by1%
  \repeat%
  }\vss }}}}%
}
\makeatother

\newcommand{\udt}[3]{#1^{#2}_{\phantom{#2}#3}}

\newcommand{\dut}[3]{#1_{#2}^{\phantom{#2}#3}}

\title{\boldmath 
Confronting quantum-corrected teleparallel cosmology with observations
}

\author[a]{Reginald Christian Bernardo,}
\author[a]{Che-Yu Chen,}
\author[b,c]{Jackson Levi Said,}
\author[d,e]{Yu-Hsien Kung}

\affiliation[a]{Institute of Physics, Academia Sinica, Taipei 11529, Taiwan}
\affiliation[b]{Institute of Space Sciences and Astronomy, University of Malta, Malta, MSD 2080}
\affiliation[c]{Department of Physics, University of Malta, Malta, MSD 2080}
\affiliation[d]{Physics Division, National Center for Theoretical Sciences, Taipei 10617, Taiwan}
\affiliation[e]{Department of Physics and Center for Theoretical Sciences, National Taiwan University, Taipei 10617, Taiwan}

\emailAdd{reginaldchristianbernardo@gmail.com}
\emailAdd{b97202056@gmail.com}
\emailAdd{jackson.said@um.edu.mt}
\emailAdd{r06222010@g.ntu.edu.tw}

\abstract{
\textbf{
It has been shown that at the semi-classical order, gravitational theories with quantum fluctuations can be effectively recast as modified theories of gravity with non-minimal gravity-matter couplings. We proceed from an observational perspective and see whether such quantum fluctuations can leave imprints on the late Universe. Within the teleparallel formulation, we investigate a representative model in this general class of modified gravitational theories inlaid with quantum fluctuations, and determine the cosmological parameters by using compiled late-time data sets. Furthermore, we assess the statistical significance of such quantum corrections compared to the standard cosmological model. The results mildly favor the inclusion of quantum corrections with a negative density parameter supporting a phantom-like dark energy. This edge is not sufficient to rule out either models but it supports the consideration of quantum corrections in a cosmological setting.
}
}

\begin{document}

\maketitle
\flushbottom

\section{Introduction}
\label{sec:intro}

The general theory of relativity (GR) enjoys its status as one of the most successful physical theories for good reasons. GR has passed astrophysical tests with unmatched precision \cite{Will:2014kxa} and even recent gravitational wave astronomy constraints \cite{Abbott:2016blz, TheLIGOScientific:2016src, PhysRevD.100.104036} such as the stringent bound on the gravitational wave speed \cite{TheLIGOScientific:2017qsa} which is in line with its GR predicted value. On cosmological scales, its parametrically simplest extension, the $\Lambda$CDM model, shares the same popularity and success \cite{SupernovaCosmologyProject:1998vns, SupernovaSearchTeam:1998fmf, WMAP:2003elm, Scolnic:2017caz, Aghanim:2018eyx}. However, GR is also undeniably ultraviolet incomplete \cite{Deser:1974cy, Deser:1974zzd, Hawking:1988wm} while at cosmological distances, the $\Lambda$CDM model can only rely on a theoretically-flawed cosmological constant to explain the current accelerating phase of cosmic expansion \cite{RevModPhys.61.1}. These, among other fundamental and observational considerations \cite{Malquarti:2003hn, Verde:2019ivm, Riess:2020fzl}, remind us to keep an open mind to the plethora of models \cite{Clifton:2011jh, Kase:2018aps, Kobayashi:2019hrl, Bahamonde:2021gfp,CANTATA:2021ktz} other than GR, and to the extension of the $\Lambda$CDM model that may better explain the dark Universe.

With only minor hints, the natural way forward theoretically is to consider fields and higher order operators that introduce departures to standard gravity at cosmological scales. The resulting modified gravity models can be then constrained using observations, leading to a selection of viable, competitive gravitational theories \cite{Peirone:2019aua, Frusciante:2019puu, Aoki:2020oqc, Anagnostopoulos:2021ydo, Atayde:2021pgb}. However, where such modified gravity terms originate is often a question that is glossed over in return of progress in other directions. We instead find motivations for these corrections by virtue of the quantum mechanical nature of the fields.

Quantum fluctuations are ubiquitous where there is matter and gravity. It is therefore important to address how such phenomena may take part in the standard model and how we are able to measure them. Progress on this was reported in Refs. \cite{Dzhunushaliev:2013nea, Dzhunushaliev:2015mva} and, more recently, in Ref. \cite{Chen:2021oal} where it was shown that semi-classically, such quantum fluctuations generally appear as modified gravity terms in the gravitational field equations. The key idea implemented in these papers is to decompose the fundamental gravitational field into a classical part and a quantum fluctuating part, may it be the metric in a curvature-based formulation or the tetrad in a teleparallel formulation. This creative step leads to a further realization that quantum fluctuations always source non-minimal couplings between matter and gravity, hinting at particle creation processes taking place through the non-conservation of the matter fields' stress-energy tensor. Coincidentally, these features are also quite generic in modified gravity theories that are often considered as an alternative to explain the late-time cosmic acceleration. This strengthens the motivation for modified gravity corrections that originate from a quantum nature. Most importantly, for our purposes, it allows us to constrain these quantum corrections through their link with modified gravity.

Building on the above formalism, the cosmological consequences arising from quantum fluctuations have been studied \cite{Yang:2015jla, Liu:2016qfx, Haghani:2021iqe}. However, to the best of our knowledge, these modified gravity generated by quantum fluctuations remain yet to be tested with current observational data. For this reason, we consider a theory (Eq. (\ref{eq:novel_lagrangian})) featuring the aforementioned aspects of quantum fluctuations as a representative of semi-classical, modified gravity models. This belongs to a class of general $f(T,B,\mathcal{T})$ gravity and was singled out in Ref. \cite{Chen:2021oal} as novel due to the absence of an equivalent Riemannian formulation. We seek to constrain the modified gravity corrections and non-minimal matter couplings in this theory using cosmological observations.

The outline of this work is as follows. We first briefly review the concept of teleparallel gravity and its equivalent description of GR, namely, the teleparallel equivalent of general relativity (Section \ref{sec:tg_quant}). Then, in the same section, we introduce the effective model that includes quantum corrections to classical gravity, which leads to the modified Friedmann equations. We then proceed to test this model with compiled late-time data sets, and present our constraints on the cosmological parameters as well as our fair assessment of the statistical significance of the results (Section \ref{sec:statistical_analysis}). We finally summarize our core results and pave some future directions on this subject (Section \ref{sec:conclusions}). Our computations are transparently communicated as jupyter notebooks and can be downloaded from \href{https://reggiebernardo.weebly.com/research.html}{GitHub}.

\section{Teleparallel gravity as generated from quantum fluctuations}
\label{sec:tg_quant}

In this section, we will first briefly review the teleparallel formulation of gravity, and how the teleparallel equivalent of general relativity (TEGR) is constructed \cite{Bahamonde:2021gfp}. After that, we will review the result of Ref.~\cite{Chen:2021oal}, in which the authors proposed an effective theory in the teleparallel formulation to describe how quantum corrections may manifest in the classical gravitational theory, in the semi-classical regimes. We will use this effective model as our main consideration to implement the cosmological tests later.
 
Einstein's GR is constructed in the Riemannian formulation, in which gravity is described solely by the curvature. In this formulation, it is the spacetime metric $g_{\mu\nu}$ that plays a fundamental role in the theory, as its dynamical variable. In addition, the connection that defines how a covariant object is parallel transported is defined through the Levi-Civita one, giving rise to the absence of torsion and non-metricity, which could be generically non-zero in general metric-affine theories \cite{Bahamonde:2021gfp}. However, GR can actually be formulated in other approaches, such as the teleparallel formulation, i.e., TEGR. In the teleparallel formulation, the affine connection is assumed in a way that the non-metricity and curvature identically vanish, and gravity is solely defined by torsion. Also, the fundamental field is the tetrad $\udt{e}{a}{\mu}$, which is an orthonormal basis that maps spacetime coordinates $x^\mu$ to the coordinates $x^a$ on the tangent-space. In terms of the tetrad, the spacetime metric can be expressed as follows
\begin{align}
    g_{\mu\nu} = \eta_{ab} \udt{e}{a}{\mu} \udt{e}{b}{\nu} \,,
\end{align}
where $\eta_{ab} = \text{diag}(1,-1,-1,-1)$. The tetrads observe the normalization conditions $\udt{e}{a}{\mu}\dut{e}{b}{\mu} = \delta_b^a$ and $\udt{e}{a}{\mu}\dut{e}{a}{\nu} = \delta_{\mu}^{\nu}$, and where $\dut{e}{b}{\mu}$ is the inverse tetrad. Note that the Greek indices represent spacetime coordinates, and the Latin indices stand for tangent-space coordinates.

In fact, in order to formulate the theory in a covariant way, one has to introduce another independent object, the spin connection, along with the tetrad to define the affine connection \cite{Bahamonde:2021srr,Bahamonde:2020bbc}. However, it turns out that the spin connection is purely inertial and one can simply choose the Weitzenb\"ock gauge in which the spin connection is identically zero to do the calculations. However, one must be careful that the antisymmetric field equations in this setting identically vanish since these are the equations of motion associated with the spin connection \cite{Bahamonde:2021gfp}. In the Weitzenb\"ock gauge, the affine connection is defined via $\Gamma^{\lambda(W)}_{\nu\mu} \equiv \dut{e}{a}{\lambda} \partial_\nu \udt{e}{a}{\mu}$, and one can define the torsion as
\begin{align}
\label{torsiontensor}
    {T^\lambda}_{\nu\mu} = \Gamma^{\lambda(W)}_{\nu\mu} - \Gamma^{\lambda(W)}_{\mu\nu} = \dut{e}{a}{\lambda} \left(\partial_\nu \udt{e}{a}{\mu} - \partial_\mu \udt{e}{a}{\nu}\right) \,.
\end{align}
It can be shown that the curvature tensors defined by this affine connection are identically zero. 

The Lagrangian density of TEGR is constructed by the torsion scalar $T$ and can be expressed as
\begin{align}
\label{Ltegr}
    \mathcal{L} = \frac{e}{2} T + e \mathcal{L}_m \,,
\end{align}
where $e = \text{det} ( \udt{e}{a}{\mu}) = \sqrt{-g}$ is the tetrad determinant, $\mathcal{L}_m$ is the matter Lagrangian, and the torsion scalar is defined by $T\equiv T_{\rho\mu\nu} S^{\rho\mu\nu}$, where the superpotential $ S^{\rho\mu\nu}$ is defined as \cite{Aldrovandi:2013wha}
\begin{align}
    S^{\rho\mu\nu} &\equiv \frac{1}{2} \left[ \frac{1}{2}\left(T^{\rho\mu\nu}+T^{\nu\mu\rho}-T^{\mu\nu\rho}\right) - g^{\rho\nu} {T^{\sigma\mu}}_\sigma + g^{\rho\mu} {T^{\sigma\nu}}_\sigma \right]
    \,.
\end{align}
Note that we will work with the unit $8\pi G/c^4=1$ throughout this paper.

The equivalence between the TEGR action (\ref{Ltegr}) and the Einstein-Hilbert action is essentially based on the fact that the torsion scalar $T$ differs from the Ricci scalar defined via the Levi-Civita connection by a total derivative term. More explicitly, one can show that
\begin{align}
\label{RQ}
    R = -T + B \,,
\end{align}
where $B=- 2 \partial_\mu \left( e {T^{\nu\mu}}_\nu \right)/e$ is a boundary term \cite{Farrugia:2020fcu,Franco:2020lxx,Caruana:2020szx}. The boundary term does not contribute to the variation, and therefore, the equations of motion in TEGR are identical to the standard Einstein equations obtained in the Riemannian formulation.

By varying the Lagrangian (\ref{Ltegr}) with respect to the tetrad $\udt{e}{a}{\mu}$, the field equation of TEGR can be obtained as
\begin{align}
    4\partial_\sigma \left( e {S_a}^{\mu\sigma} \right) - 4e {S_c}^{\nu\mu} {T^c}_{\nu a} + e \dut{e}{a}{\mu} T - 2  e \mathcal{T}_a^\mu = 0 \,,\label{classiceom}
\end{align}
where
\begin{align}
    \mathcal{T}_a^\mu = -\frac{1}{e} \frac{\delta ( e L_m )}{\delta \udt{e}{a}{\mu}} 
\end{align}
is the energy-momentum tensor in TEGR. As we have just mentioned, since the Ricci scalar in the Riemannian formulation, i.e., $R$, and the torsion scalar $T$ differ by a total derivative term $B$, it can be explicitly shown that the equation of motion (\ref{classiceom}) is dynamically equivalent to the GR field equations in their classical limit. That is why the theory is dubbed ``TEGR".

Let us include quantum corrections into the story. In Refs. \cite{Dzhunushaliev:2013nea, Dzhunushaliev:2015mva}, an effective approach for incorporating quantum corrections into the classical gravitational theory in the Riemannian formulation was proposed. This approach is based on the assumptions that all the classical entities in the theory are replaced by quantum operators, and, in the semi-classical regimes, the metric operator can be decomposed into a classical part and a quantum fluctuating part. The quantum effects in the resulting effective theories generically give rise to non-minimal couplings between matter and geometry. Very recently, this approach was applied to the teleparallel formulation \cite{Chen:2021oal} and the resulting effective theories are generically featured by non-minimal torsion-matter couplings.

Let us briefly review the results in Ref.~\cite{Chen:2021oal}. As we have just mentioned, we start with the assumption that the tetrad is quantum in nature and it can be expressed as a quantum operator $\widehat{\udt{e}{a}{\mu}}$. Then, the tetrad operator is decomposed into a classical part and a quantum part:
\begin{equation}
    \widehat{\udt{e}{a}{\mu}}=\udt{e}{a}{\mu}+\widehat{\delta \udt{e}{a}{\mu}}\,.
\end{equation}
In the semi-classical regimes, we neglect higher-order contributions from the quantum part of the tetrad operator, and expand the operator version of the TEGR Lagrangian as
\begin{align}
    \widehat{\mathcal{L}}&=\frac{1}{2}\widehat{e}\widehat{T}+\widehat{e}\widehat{\mathcal{L}_m}\nonumber\\
    &\approx\frac{1}{2}eT+e\mathcal{L}_m+\frac{1}{2}\frac{\delta\left(eT\right)}{\delta \udt{e}{a}{\rho}}\widehat{\delta \udt{e}{a}{\rho}}+\frac{\delta\left(e\mathcal{L}_m\right)}{\delta \udt{e}{a}{\rho}}\widehat{\delta \udt{e}{a}{\rho}}\,.
\end{align}
Note that the coefficients in the third and the fourth terms in the second line are given by the variations of the classical TEGR Lagrangian with respect to the tetrad $\udt{e}{a}{\rho}$, giving rise to the TEGR equation of motion (\ref{classiceom}). Therefore, we obtain
\begin{align}
    \widehat{\mathcal{L}}=&\frac{1}{2}eT+e\mathcal{L}_m+ \frac{e}{2} \left[ \frac{4}{e} \partial_\sigma \left( e {S_a}^{\rho\sigma} \right)
    - 4{S_c}^{\nu\rho}  {T^c}_{\nu \lambda} e_a^\lambda + \dut{e}{a}{\rho} T - 2  \mathcal{T}_a^\rho \right] \widehat{\delta \udt{e}{a}{\rho}}\,.\label{langrangiannoaverage}
\end{align}
To proceed, we adopt the idea of Refs. \cite{Dzhunushaliev:2013nea, Dzhunushaliev:2015mva} and assume that the average of the quantum fluctuating part of the tetrad is not zero, and is constructed by a tetrad-like field:
\begin{equation}
    \langle\widehat{\delta \udt{e}{a}{\rho}}\rangle=\udt{Q}{a}{\rho}\ne0\,.
\end{equation}
In this regard, the effective Lagrangian can be obtained by taking an average of the Lagrangian operator (\ref{langrangiannoaverage}):
\begin{equation}
    \mathcal{L}_\text{eff}\equiv\langle\widehat{\mathcal{L}}\rangle=\frac{1}{2}eT+e\mathcal{L}_m+ \frac{e}{2} \left[ \frac{4}{e} \partial_\sigma \left( e {S_a}^{\rho\sigma} \right)
    - 4{S_c}^{\nu\rho}  {T^c}_{\nu \lambda} \dut{e}{a}{\lambda} + \dut{e}{a}{\rho} T - 2  \mathcal{T}_a^\rho \right] \udt{Q}{a}{\rho}\,.
\end{equation}

At this point, the quantum tetrad $\udt{Q}{a}{\rho}$ remains arbitrary. However, it can be chosen by considering the following natural guidance. First, the quantum tetrad is built solely by combinations of classical quantities. Second, $\udt{Q}{a}{\rho}$ mush be chosen in a way that it approaches zero in the absence of gravity. In Ref.~\cite{Chen:2021oal}, an effective theory was obtained by explicitly choosing
\begin{equation}
    \udt{Q}{a}{\rho}=\alpha T \udt{e}{a}{\mu}\,,
\end{equation}
where $\alpha$ is a constant quantifying the amount of quantum corrections. The effective Lagrangian can then be expressed as
\begin{equation}
\label{eq:novel_lagrangian}
\mathcal{L}_\text{eff} = \dfrac{1}{2} \left[ T + 2 \alpha T \left( T - B \right) + 2 \mathcal{L}_m - 2 \alpha T \mathcal{T} \right]e
\end{equation}
where $\mathcal{T}$ is the trace of the matter stress-energy tensor. This theory can be singled out as it has no Riemannian counterpart. In addition, the effective Lagrangian (\ref{eq:novel_lagrangian}) belongs to a general $f(T,B,\mathcal{T})$ gravity in which the trace of the energy-momentum tensor $\mathcal{T}$ is non-minimally coupled with the torsion scalar $T$ and the boundary $B$. The equation of motion of the general $f(T,B,\mathcal{T})$ gravity, whose action is given by
\begin{equation}
    \mathcal{S}=\frac{1}{2}\int d^4xef(T,B,\mathcal{T})+\mathcal{S}_m\,,
\end{equation}
has been obtained in Ref.~\cite{Chen:2021oal} and it reads:
\begin{align}
\frac{1}{2}f\dut{e}{a}{\nu} &+ 2f_T{T^\rho}_{\mu a}{S_\rho}^{\nu\mu} -\frac{2}{e}\partial_\mu\left(ef_T{S_{a}}^{\mu\nu}\right)+\dut{e}{a}{\rho}\nabla^\nu\nabla_\rho f_B \nonumber \\
& \phantom{ggggggggg} -\dut{e}{a}{\nu}\Box f_B-\frac{1}{2}Bf_B\dut{e}{a}{\nu}-2\left(\partial_\mu f_B\right){S_{a}}^{\mu\nu}
= -\frac{f_\mathcal{T}}{2}\left(g^{\alpha\beta}\frac{\partial\mathcal{T}_{\alpha\beta}}{\partial \udt{e}{a}{\nu}}-2\mathcal{T}_a^\nu\right)+\mathcal{T}_{a}^\nu\,,\label{eomgeral}
\end{align}
where $f_{T,B,\mathcal{T}}$ denote the partial derivatives of $f$ with respect to its argument. On the above equation, $\nabla_\mu$ denotes the covariant derivative defined in terms of the Levi-Civita
connection. The effective theory given by the Lagrangian (\ref{eq:novel_lagrangian}) simply corresponds to the following functional $f$
\begin{equation}
    f(T,B,\mathcal{T})=T+2\alpha T\left(T-B\right)-2\alpha T\mathcal{T}\,.
\end{equation}
It should be mentioned that the theory is featured by the non-minimal torsion-matter couplings, i.e., the $T\mathcal{T}$ coupling term, and the higher-order derivative terms contributed by the boundary term $B$.

In order to investigate the possibility of constraining the model using cosmological data, we consider the Friedmann–Lemaître–Robertson–Walker (FLRW) metric, in which the evolution of a homogeneous and isotropic Universe is solely described by its scale factor $a(t)$ as a function of the cosmic time $t$. Then, we assume that the Universe is dominated by perfect fluids with \textit{bare} energy density $\rho$ and pressure $P$. By using the equation of motion (\ref{eomgeral}), we arrive at the modified Friedmann equations
\begin{equation}
\label{eq:FeqM}
3H^2 = \rho - 6 \alpha H^2 \left( \rho + 5 P + 18 H^2 \right) \,,
\end{equation}
\begin{equation}
\label{eq:PeqM}
-2 \dot{H} - 3 H^2 =  P + 2 \alpha \left[ 54 H^4 + 2 \dot{H} \left( -\rho + 3 P \right) + 3 H^2 \left( - \rho + 3 P + 24 \dot{H} \right) + 2 H \left( - \dot{\rho} + 3 \dot{P} \right) \right] \,,
\end{equation}
and a modified conservation equation
\begin{equation}
\label{eq:Meq}
\dot{\rho} + 3H \left( \rho + P \right) = -6\alpha H \left[ -2 \left( 2 \dot{H} + 3 H^2 \right) \left( \rho + P \right) + H \left( - 3 \dot{\rho} + \dot{P} \right) \right] \,,
\end{equation}
where $H\equiv\dot{a}/a$ is the Hubble rate and the dot denotes the derivative with respect to the cosmic time $t$. An alternative derivation is also shown in Appendix \ref{sec:alt_derivation} utilizing an exact mapping of the bare fluid and scalar field quantities. Clearly, the above equations describe a non-minimally coupled fluid. The standard cosmological model can be recovered in the limit $\alpha \rightarrow 0$. It is also useful to note that the above modified field equations retain the property that only two of the three equations are independent, e.g., differentiating Eq. (\ref{eq:FeqM}) and using Eq. (\ref{eq:PeqM}) leads to Eq. (\ref{eq:Meq}). Furthermore, for multiple fluids, each on its own trajectory, it can be shown that their bare energy densities and pressures sum up in the Friedmann equations. These comments refer to a particularly notable quirk of these modified equations, to be described next, and also play a role in the numerical integration.

For our purposes, the most important thing to be said about the modified field equations is that matter no longer follows its standard path. This is expected considering that the theory falls under $f\left(T, B, \mathcal{T}\right)$. But to be clear about this, let us consider the cold, pressureless fluid $P = 0$. In this case, the field equation of the fluid becomes
\begin{equation}
\label{eq:Meq_pressureless}
\dot{\rho} + 3H \rho = -6\alpha H \left[ -2 \left( 2 \dot{H} + 3 H^2 \right) \rho - 3 H \dot{\rho} \right] \,.
\end{equation}
This can be solved in a semi-closed form as an integral expression; however, the result is obviously no longer as pretty as the usual $\alpha \rightarrow 0$ solution $\rho \sim 1/a^3$. Nonetheless, vacuum energy $\left(P_\Lambda = -\rho_\Lambda\right)$ turns out to still be solvable by $\rho = \Lambda$ where $\Lambda$ is a constant. This simplifies the analysis of the late Universe to practically a numerical integration of the coupled system for the Hubble function and the pressureless fluid.

In what follows, we shall constrain the quantum corrections with compiled late-time cosmological data sets. In doing so, we express the independent variable in terms of the redshift $z = \left(1/a\right) - 1$ and write down $\alpha = q/H_0^2$, where $H_0$ is the Hubble parameter at $z = 0$, such that $q$ stands for a dimensionless quantum parameter. This is notably convenient numerically as $q$ is dimensionless. The Friedmann constraint at $z = 0$ $\left( \Omega_{m0} + \Omega_\Lambda + \Omega_{q0} = 1 \right) $ determines the density parameter of the quantum corrections $\Omega_{q0}$. We will find constraints on this quantity in the next section. 

\section{Observational constraint on the quantum contribution}
\label{sec:statistical_analysis}

In Section \ref{subsec:late_time}, we briefly review the data sets to be considered in the analysis. Then, we setup the dynamical system for the numerical implementation in Section \ref{subsec:dynamical_system}, and proceed to our main results in Section \ref{subsec:bayesian} where we find best fit values for the quantum parameter. We refer to the model as TG/quant for brevity.

\subsection{Late-time observations}
\label{subsec:late_time}

We consider compiled data sets from cosmic chronometers (CC), baryon acoustic oscillations (BAO), and supernovae (SNe). 

As the base of our analysis, we take into account Hubble function data $H(Z)$, where uppercase $Z$ (or $Z'$) refers to the discrete, observation redshifts in contrast with the general redshifts $z$. The first part (CC) of this comes from a differential aging method involving passively evolving galaxies and consists of 31 measurements \cite{Moresco:2016mzx, Moresco:2015cya, 2014RAA....14.1221Z, 2010JCAP...02..008S, 2012JCAP...08..006M,Ratsimbazafy:2017vga}. This relies on measurements of redshifts and ages of adjacent galaxies in such a way that the Hubble function can be approximated as $H(z) = \dot{a}/a \sim (\Delta z/ \Delta t)/(1 + z)$. The second part comes from line-of-sight baryon acoustic oscillations (BAO) to make up 26 more data points  \cite{2012MNRAS.425..405B, Chuang:2013hya, BOSS:2013igd, BOSS:2014hwf, Bautista:2017zgn} which are not correlated. BAO are fluctuations in the baryon-photon plasma during the early drag epoch which left out observational imprints on the sky often referred to as standard rulers. Naturally, this is because the BAO capture the comoving sound horizon radius during baryon drag.

We also consider the 1048 SNe type Ia observations of Pantheon spanning the redshifts $0.01 < z < 2.3$ \cite{Scolnic:2017caz} as well as the SNe absolute magnitude. This is in terms of the SNe apparent magnitudes at their brightest and comes with a full covariance matrix correlating the measurements at various redshifts. When using the Pantheon data, we additionally take into account the SH0ES prior for the SNe absolute magnitude ($M = -19.22 \pm 0.04$) wherein the SNe are calibrated using cepheids \cite{Riess:2019cxk}.

\subsection{The modified Friedmann equations}
\label{subsec:dynamical_system}

In the matter sector, we consider a cosmological constant $\Lambda$ together with a pressureless perfect fluid with an energy density $\rho(z)$ representing the contribution from baryons and cold dark matter. In this case, the Friedmann equations and the fluid equation can respectively be written as
\begin{equation}
\label{eq:friedmann}
6 \alpha  H^2 \left(18 H^2-4 \Lambda +\rho \right)+3 H^2-\Lambda -\rho =0 \,,
\end{equation}
\begin{equation}
\label{eq:pressure}
\begin{split}
144 \alpha  H^3 (z+1) H'+\Lambda =
H \bigg[ 108 \alpha  H^3 & +2 (z+1) H' (8 \alpha  \Lambda +2 \alpha  \rho -1) \\
& + H \left(-24 \alpha  \Lambda -6 \alpha  \rho +4 \alpha  (z+1) \rho '+3\right) \bigg] \,,
\end{split}
\end{equation}
and
\begin{equation}
\label{eq:mkg}
(z+1) \left(18 \alpha  H^2-1\right) \rho '+3 \rho  \left(4 \alpha  H \left(2 (z+1) H'-3 H\right)+1\right) = 0 \,,
\end{equation}
where the primes now indicate differentiation with respect to the redshift $z$. Naively, this may be read as a two-dimensional dynamical system with the dependent variables $\left( H, \rho \right)$ whose dynamics is determined by Eqs. (\ref{eq:pressure}) and (\ref{eq:mkg}). But, the Friedmann constraint (\ref{eq:friedmann}) shows that the overall system should be inherently one-dimensional, which turns to our advantage.

We simplify the numerical analysis further by defining dimensionless constants $(q, \lambda, \Omega_{m0})$, utilizing the present Hubble expansion rate $H_0$ as a length scale,
\begin{eqnarray}
q &=& \alpha H_0^2 \\
\Omega_\Lambda &=& \Lambda/(3 H_0^2) \\
\Omega_{m0} &=& \rho(z = 0)/(3 H_0^2) \,,
\end{eqnarray}
which, in terms of which the Friedmann constraint at $z = 0$, leads to the relation
\begin{equation}
\label{eq:friedmann0rel}
1 = \Omega_\Lambda + \Omega_{m0} + q \left[ 24 \Omega_\Lambda - 6 \left( 6 + \Omega_{m0} \right) \right] \,.
\end{equation}
Clearly, Eq. (\ref{eq:friedmann0rel}) permits the identification of $\Omega_{m0}$ and $\Omega_\Lambda$ as the \textit{bare} contributions to the total energy fraction while the quantum corrections
\begin{equation}
\label{eq:oq0}
\Omega_{q0} = q \left[ 24 \Omega_\Lambda - 6 \left( 6 + \Omega_{m0} \right) \right]
\end{equation}
contribute to the present expansion. We also define the dimensionless Hubble function
\begin{equation}
\label{eq:ydef}
E(z) = H(z)/H_0\,,
\end{equation}
using which the dynamical system can be explicitly written in a dimensionless form as
\begin{equation}
\label{eq:master_eq}
E' = \dfrac{3 \left(18 q E^2 \left(4 q E^2-1\right)+1\right) \left(-\Omega_\Lambda +36 q E^4+E^2 (1-24 \Omega_\Lambda  q)\right)}{2 E (z+1) \left(-18 \Omega_\Lambda  q+6 q E^2 \left(126 \Omega_\Lambda  q+24 q E^2 \left(-12 \Omega_\Lambda  q+45 q E^2-13\right)+7\right)+1\right)} \,.
\end{equation}
Particularly, this can be obtained by eliminating $\rho(z)$ in Eq. (\ref{eq:mkg}) using the Friedmann constraint (\ref{eq:friedmann}).

The numerical integration can then be started at $z = 0$ with $E_0 = E(z = 0) = 1$ provided a choice of the theory parameters which we conveniently choose to be $H_0$, $\Omega_{m0}$, and $\varepsilon = -\Omega_{q0}$. Therefore, in principle, we shall be sampling over the bare matter fraction $\Omega_{m0}$ and the quantum corrections $\Omega_{q0}$ while $\Omega_\Lambda$ is determined by the Friedmann constraint. The $\Lambda$CDM limit can be obtained by simply evaluating $\Omega_{q0} = 0$ in this parameter space.

\subsection{Bayesian analysis of the model with late-time data}
\label{subsec:bayesian}

To obtain the best fit estimates of the cosmological parameters, we sample over the parameter space via a Markov chain Monte-Carlo routine. This determines the joint posterior of the cosmological parameters where we keep the likelihood of each of the data set to be their $\chi^2$ values, i.e., for the base Hubble data (CC + BAO),
\begin{equation}
\label{eq:chi2_Hz}
    \chi^2_{H(Z)} = \sum_Z \left( \dfrac{ H_\text{obs}(Z) - H_\text{model}(Z) }{\sigma_{H, \text{obs}}(Z)} \right)^2 \,,
\end{equation}
and for the supernovae,
\begin{equation}
\label{eq:chi2_sne}
    \chi^2_\text{SNe} = \sum_{Z,Z'} \left( m_\text{obs}(Z) - m_\text{model}(Z) \right) C^{-1}(Z, Z') \left( m_\text{obs}(Z') - m_\text{model}(Z') \right) \,.
\end{equation}
In Eqs. (\ref{eq:chi2_Hz}) and (\ref{eq:chi2_sne}), the subscripts ``obs'' and ``model'' refer to the observation and the model values of the Hubble parameter, while $Z$ and $Z'$ are when the observations took place, $\sigma_H$ are the uncertainties in the measurements of $H(Z)$, and $m(Z)$ are the SNe apparent magnitudes with a data covariance matrix $C(Z, Z')$. We perform the sampling using the python package \href{https://cobaya.readthedocs.io/}{cobaya} \cite{2020arXiv200505290T} and keep a stringent Gelman-Rubin convergence criterion of $R - 1 = 10^{-3}$ throughout. We consider flat priors and reference values shown in Table \ref{tab:flat_priors} for the cosmological parameters while taking in the SH0ES prior for the SNe absolute magnitude ($M = -19.22 \pm 0.04$). The varying reference values ensure a better reliability of the convergence of the samples than when a single starting point is used.

\begin{table}[h!]
    \centering
    \begin{tabular}{| c | c | c |} \hline
    \phantom{ $\dfrac{1}{1}$ } Parameter \phantom{ $\dfrac{1}{1}$ } & \phantom{ $\dfrac{1}{1}$ } Prior \phantom{ $\dfrac{1}{1}$ } & \phantom{ $\dfrac{1}{1}$ } Ref \phantom{ $\dfrac{1}{1}$ } \\ \hline \hline
    \phantom{ $\dfrac{1}{1}$ } $H_0$ [km s$^{-1}$ Mpc$^{-1}$] \phantom{ $\dfrac{1}{1}$ } & \phantom{ $\dfrac{1}{1}$ } $\left( 50, 80 \right)$ \phantom{ $\dfrac{1}{1}$ } & \phantom{ $\dfrac{1}{1}$ } $\left( 68, 72 \right)$ \phantom{ $\dfrac{1}{1}$ } \\ \hline
    \phantom{ $\dfrac{1}{1}$ } $\Omega_{m0}$ \phantom{ $\dfrac{1}{1}$ } & \phantom{ $\dfrac{1}{1}$ } $\left( 0, 1 \right)$ \phantom{ $\dfrac{1}{1}$ } & \phantom{ $\dfrac{1}{1}$ } $\left( 0.25, 0.35 \right)$ \phantom{ $\dfrac{1}{1}$ } \\ \hline
    \phantom{ $\dfrac{1}{1}$ } $\varepsilon = - \Omega_{q0}$ \phantom{ $\dfrac{1}{1}$ } & \phantom{ $\dfrac{1}{1}$ } $\left( -0.07, 0.07 \right)$ \phantom{ $\dfrac{1}{1}$ } & \phantom{ $\dfrac{1}{1}$ } $\left( -0.01, 0.01 \right)$ \phantom{ $\dfrac{1}{1}$ } \\ \hline \hline
    \end{tabular}
    \caption{Priors and reference values (Ref) used for the sampling of the cosmological parameters $\left( H_0, \Omega_{m0}, \varepsilon \right)$.}
    \label{tab:flat_priors}
\end{table}

\begin{figure}[h!]
\center
\includegraphics[width = 0.65 \textwidth]{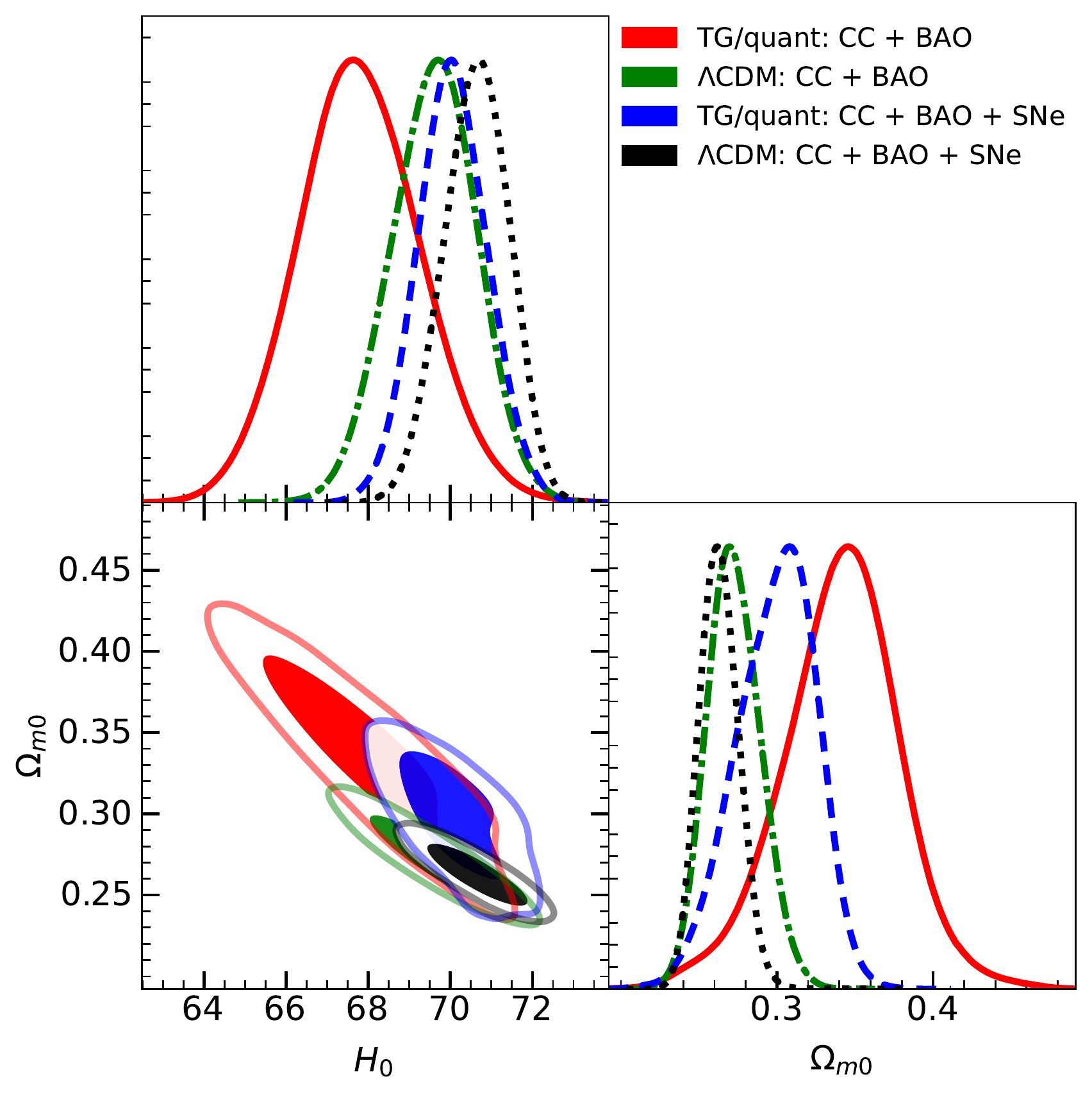}
\caption{Posteriors of the Hubble parameter and the matter fraction at $z = 0$ for TG/quant and $\Lambda$CDM for the lone Hubble data (CC + BAO) and ones including supernovae (SNe).}
\label{fig:post_H0_om0_All}
\end{figure}

Figure \ref{fig:post_H0_om0_All} shows the posteriors of the Hubble parameter $H_0$ and the matter fraction $\Omega_{m0}$ at $z = 0$ for both TG/quant and $\Lambda$CDM models.

\begin{figure}[h!]
\center
\includegraphics[width = 0.65 \textwidth]{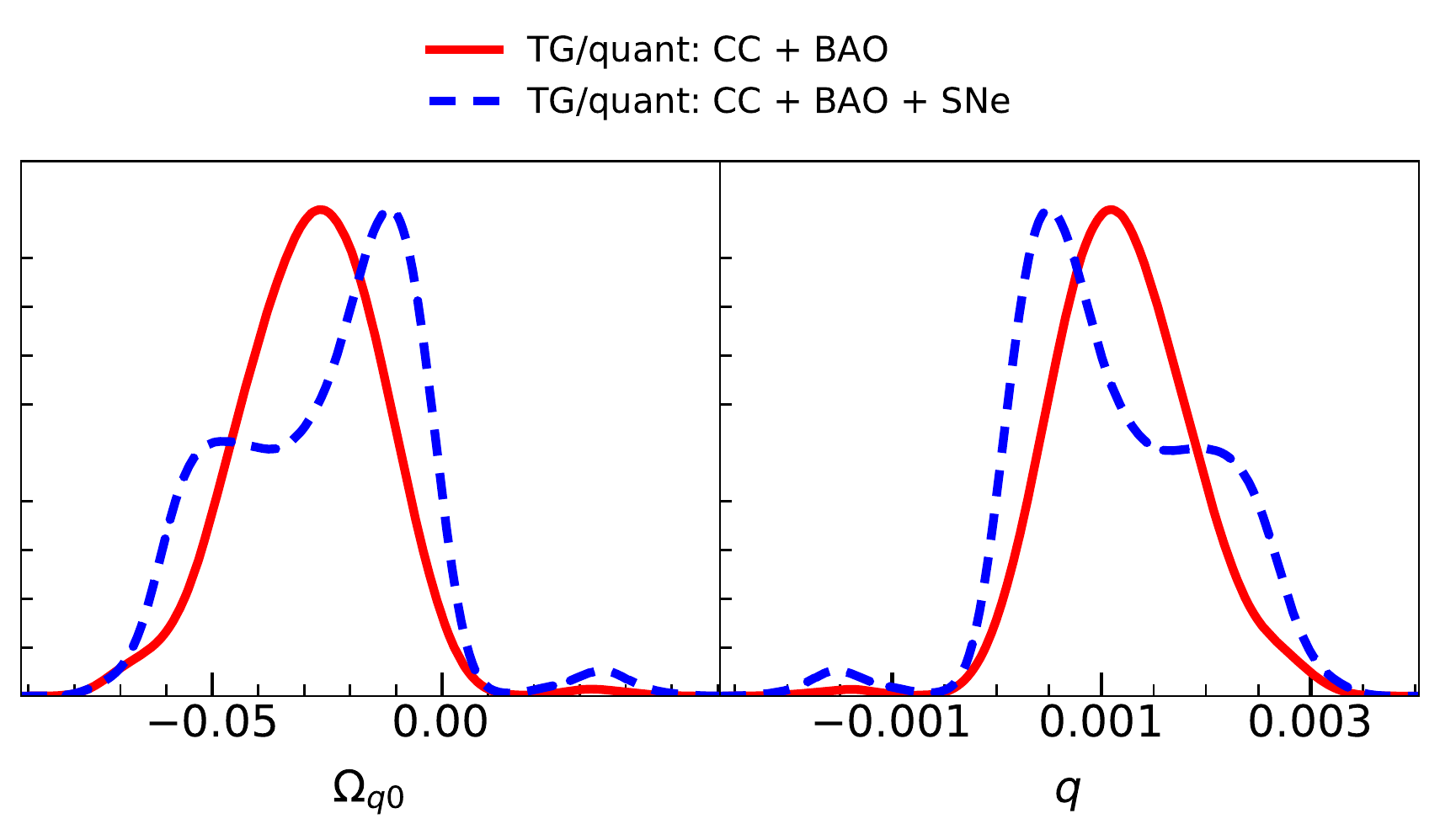}
\caption{One-dimensional posteriors of the parameters $\Omega_{q0}$ and $q = \alpha H_0^2$ indicating that the data tends to slightly prefer nonzero quantum corrections. This holds with the lone Hubble data (CC + BAO) and ones including supernovae (SNe). Both $\Omega_{q0}$ and $q$ are included in this plot for whichever is more convenient to interpret.}
\label{fig:post_oq0_q_TG}
\end{figure}

It can be seen that when the quantum corrections are included (TG/quant), $H_0$ tends to lower values while $\Omega_{m0}$ to higher values compared to their $\Lambda$CDM counterparts. This holds with just the base Hubble data (CC + BAO) and even with supernovae observations. This is also reflected in Table \ref{tab:bestfits} where the best fit values of each relevant cosmological parameter for each model and data set are presented. On the other hand, the density parameter $\Omega_{q0}$ determining the possible contribution of quantum corrections in the late Universe is shown in Figure \ref{fig:post_oq0_q_TG}. The posterior of $q = \alpha H_0^2$ is also presented.

Figure \ref{fig:post_oq0_q_TG} reflects an overall preference for $\Omega_{q0} < 0$ for CC + BAO and CC + BAO + SNe. This translates to $q > 0$ noting that $\Omega_{q0}$ is not trivially proportional to $q$, but depends also on other parameters in the system  (\ref{eq:oq0}). Nonetheless, it is illustrative to present both since $\Omega_{q0}$ represents the fraction of quantum corrections in the late Universe (relevant for cosmologists) while $q$ denotes the parameter which directly enters the action (relevant for theorists). The best fit values (first and second moments) of $\Omega_{q0}$ are shown in Table \ref{tab:bestfits} together with $H_0$ and $\Omega_{m0}$ for the $\Lambda$CDM counterparts.

\begin{table}[h!]
\center
\caption{Marginalized statistics of the cosmological parameters $H_0$, $\Omega_{m0}$, and $\Omega_{q0}$.}
\begin{tabular}{| c | c | c | c | c |}
\hline

Model & \phantom{ $\dfrac{1}{1}$ } Data set \phantom{ $\dfrac{1}{1}$ } & \phantom{ $\dfrac{1}{1}$ } $H_0$ \phantom{ $\dfrac{1}{1}$ } & \phantom{ $\dfrac{1}{1}$ } $\Omega_{m0}$ \phantom{ $\dfrac{1}{1}$ } & \phantom{ $\dfrac{1}{1}$ } $\Omega_{q0}$ \phantom{ $\dfrac{1}{1}$ } \\ \hline \hline

\multirow{2}{*}{TG/quant}
& \phantom{ $\dfrac{1}{1}$ } CC + BAO \phantom{ $\dfrac{1}{1}$ } & $67.8 \pm 1.5$ & $0.34 \pm 0.04$ & $-0.03 \pm 0.02$ \\

& \phantom{ $\dfrac{1}{1}$ } CC + BAO + SNe \phantom{ $\dfrac{1}{1}$ } & $70.1 \pm 0.9$ & $0.30 \pm 0.03$ & $-0.03 \pm 0.02$ \\
\hline \hline

\multirow{2}{*}{$\Lambda$CDM}
& \phantom{ $\dfrac{1}{1}$ } CC + BAO \phantom{ $\dfrac{1}{1}$ } & $69.6 \pm 1.0$ & $0.27 \pm 0.02$ & $0$ \\

& \phantom{ $\dfrac{1}{1}$ } CC + BAO + SNe \phantom{ $\dfrac{1}{1}$ } & $70.6 \pm 0.8$ & $0.26 \pm 0.01$ & $0$ \\
\hline \hline
\end{tabular}
\label{tab:bestfits}
\end{table}

Table \ref{tab:bestfits} shows the interplay between the standard cosmological parameters $\left(H_0, \Omega_{m0}\right)$ and quantum corrections $\left(\Omega_{q0}\right)$. Clearly, the inclusion of quantum corrections tilts $H_0$ to lower values and $\Omega_{m0}$ to higher ones than in $\Lambda$CDM, albeit leading to broader distributions as also seen previously (Figure \ref{fig:post_H0_om0_All}). It may be useful to note that the Friedmann constraint guarantees $\Omega_{m0} + \Omega_\Lambda + \Omega_{q0} = 1$ where $\Omega_\Lambda$ corresponds to the fraction of $\Lambda$ sourcing the current phase of cosmic acceleration. This may be the reason for $\Omega_{m0}$ shifting to higher values since $\Omega_{q0}< 0$ is clearly being preferred by the observations. However, it should be noted that the shape of the posteriors of $\Omega_{q0}$ (and consequently $q$) are nonstandard particularly with the SNe data (Figure \ref{fig:post_oq0_q_TG}) and so the first two moments alone do not capture the most accurate picture of the distribution.

The addition of SNe leads to higher values of $H_0$. This behavior can be traced to the BAO data whose consideration is known to lead to lower $H_0$. It is also interesting that the quantum fluctuations lead to lower $H_0$ values in the late Universe. Understandably, we focus on the significance of quantum fluctuations to keep the narrative simple. Nonetheless, in light of the Hubble tension, this interplay between $H_0$ and $\Omega_{q0}$, and the data sets, may be useful to keep in mind in future work.

Another notable observation arising from Table \ref{tab:bestfits} is that the $H_0$ values in TG/quant alone with and without the supernovae observations are in statistical tension with each another. To be concrete, the $H_0$ measurement with CC + BAO is in a $2.5\sigma$ tension with the $H_0$ value obtained with the supernovae. Interestingly, this tension does not appear on the density parameter of the quantum fluctuations $\Omega_{q0}$ nor on the standard matter fields $\Omega_{m0}$. This may be taken to be a pathology of the model or an indication of an elementary physics requiring further examinations.

\begin{figure}[h!]
\center
	\subfigure[ ]{
		\includegraphics[width = 0.6 \textwidth]{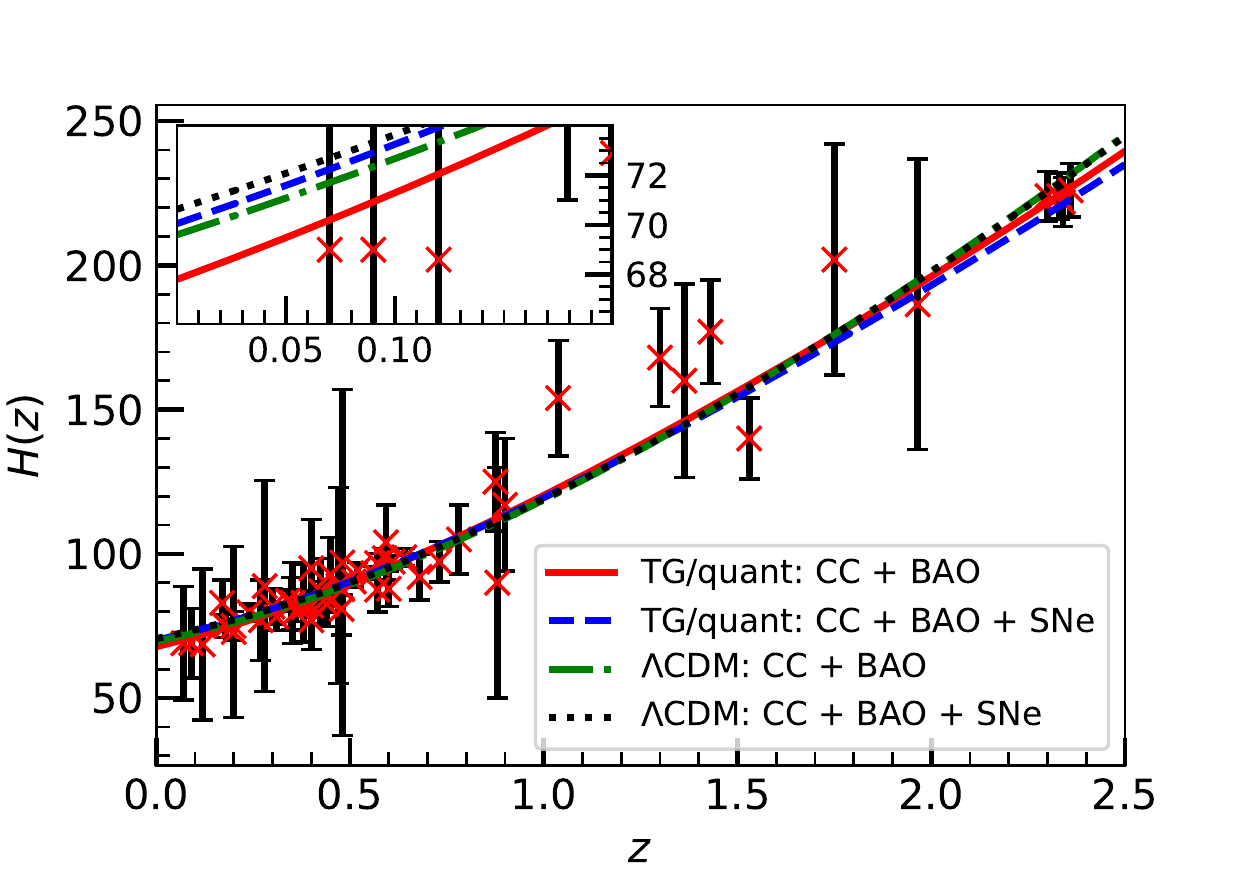}
		}
	\subfigure[ ]{
		\includegraphics[width = 0.6 \textwidth]{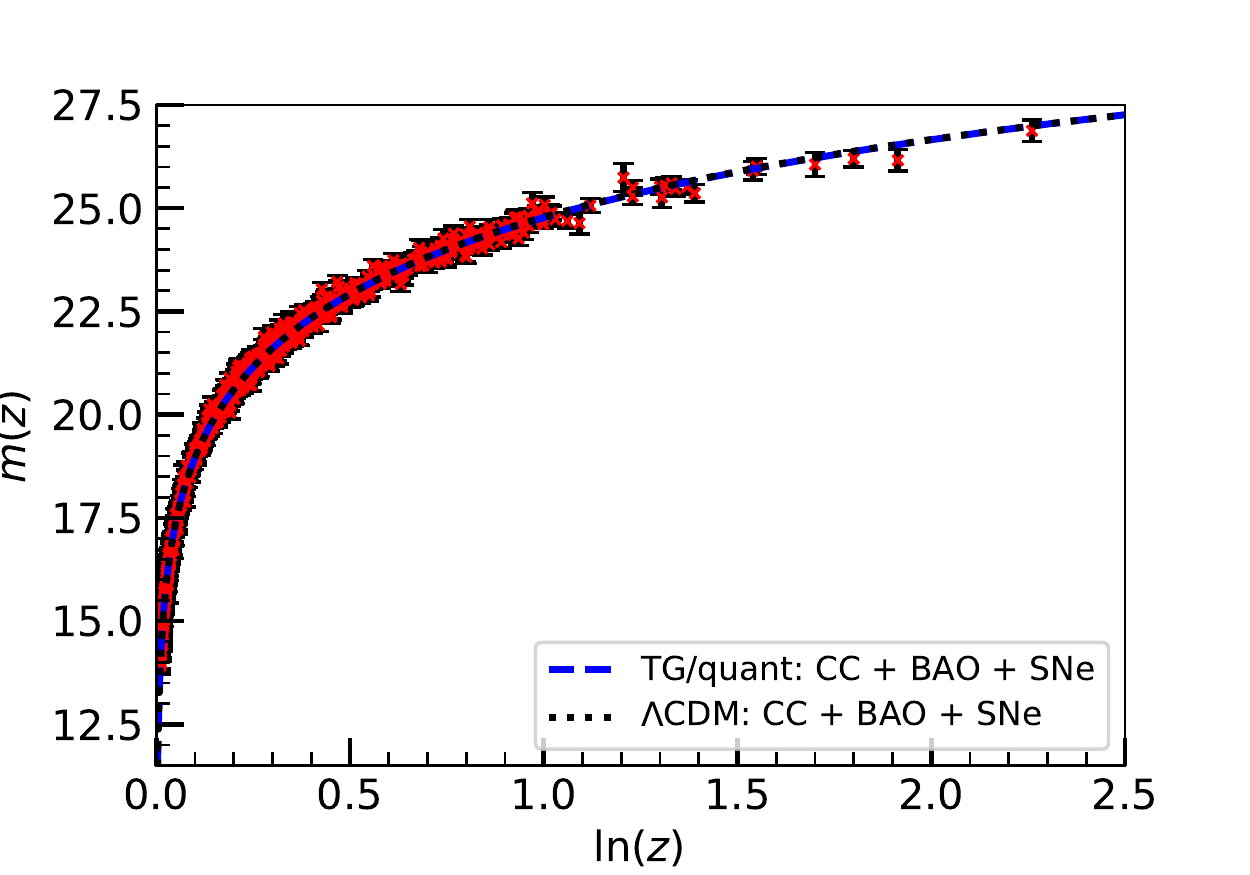}
		}
\caption{Best fit curves arising from both TG/quant and $\Lambda$CDM models with CC + BAO and CC + BAO + SNe data sets. This is shown together with plots of (a) the Hubble data and (b) the supernovae apparent magnitude. The inset in (a) shows a magnified low redshift region $z \in \left( 0, 0.2 \right)$ of the best fit curves of the Hubble function.}
\label{fig:bestfit_Hz_mz_All}
\end{figure}

The mean curves drawn from the sampling is shown in Figure \ref{fig:bestfit_Hz_mz_All} with the corresponding data sets that were used for the analysis. This comprises of the 57 Hubble data points from CC and BAO and the 1048 SNe samples from the Pantheon compilation.

It can be seen that in TG/quant, $H(z)$ for the higher redshifts tend to lower values compared to $\Lambda$CDM. This is notable considering that it is the same behavior as what was observed at the low redshifts (see $H_0$ in Table \ref{tab:bestfits}). The curves of $H(Z)$ per model can be visually distinguished as shown. On the other hand, the curves for the SNe apparent magnitudes are practically visually indistinguishable in TG/quant and $\Lambda$CDM. This can be expected due to the Pantheon samples' small uncertainties and covariance matrix for all 1048 SNe observations.

We must also assess the statistical significance of the quantum corrections. For this purpose, we consider the chi-squared ($\chi^2$), Akaike information criterion (AIC), and the Bayesian information criterion (BIC) for the TG/quant and $\Lambda$CDM models. The $\chi^2$ measures how many standard deviations away the model is from the data. On the other hand, the AIC and BIC additionally penalize complexity:
\begin{equation}
\text{AIC} = 2 k - 2 \log \left( \hat{ \mathcal{L} } \right) ,
\end{equation}
\begin{equation}
\text{BIC} = k \log \left( N \right) - 2 \log \left( \hat{ \mathcal{L} } \right) ,
\end{equation}
where $k$ is the number of sampled parameters, $N$ is the size of the data, and $\hat{ \mathcal{L} }$ is the maximum likelihood of the model. We then subtract the $\chi^2$, AIC, and BIC for TG/quant from their $\Lambda$CDM counterparts leading to $\Delta \chi^2$, $\Delta$AIC, and $\Delta$BIC. Positive values for these metrics imply statistical preference for the inclusion of quantum corrections. The results are presented in Table \ref{tab:statistics}.

\begin{table}[h!]
\center
\caption{Statistical assessment of TG/quant compared to $\Lambda$CDM in terms of the $\chi^2$, the Akaike information criterion (AIC), and the Bayesian information criterion (BIC). A preference of TG/quant corresponds to positive values.}
\begin{tabular}{| c | c | c | c | c |}
\hline

Model & \phantom{ $\dfrac{1}{1}$ } Data set \phantom{ $\dfrac{1}{1}$ } & \phantom{ $\dfrac{1}{1}$ } $\Delta \chi^2$ \phantom{ $\dfrac{1}{1}$ } & \phantom{ $\dfrac{1}{1}$ } $\Delta$AIC \phantom{ $\dfrac{1}{1}$ } & \phantom{ $\dfrac{1}{1}$ } $\Delta$BIC \phantom{ $\dfrac{1}{1}$ } \\ \hline \hline

\multirow{2}{*}{TG/quant}
& \phantom{ $\dfrac{1}{1}$ } CC + BAO \phantom{ $\dfrac{1}{1}$ } & $3.46$ & $1.46$ & $-0.59$ \\

& \phantom{ $\dfrac{1}{1}$ } CC + BAO + SNe \phantom{ $\dfrac{1}{1}$ } & $6.33$ & $4.33$ & $1.75$ \\
\hline \hline
\end{tabular}
\label{tab:statistics}
\end{table}

It can be seen that in terms of the $\chi^2$, the inclusion of quantum corrections is statistically preferred with data sets CC + BAO and CC + BAO + SNe. This is also true for the AIC which is positive for both cases. The BIC, on the other hand, slightly gives the edge to $\Lambda$CDM with the CC + BAO data set but prefers TG/quant with the added supernovae observations. Overall, when all the numbers are taken into consideration, it seems fair to conclude that the data slightly prefers the inclusion of quantum corrections generating modified teleparallel gravity and non-minimal gravity-matter couplings. This is more so true when supernovae observations are taken into account. However, it must be said that as far as the BIC is concerned, this edge towards TG/quant can still be reconsidered.

\begin{figure}[h!]
\center
\includegraphics[width = 0.6 \textwidth]{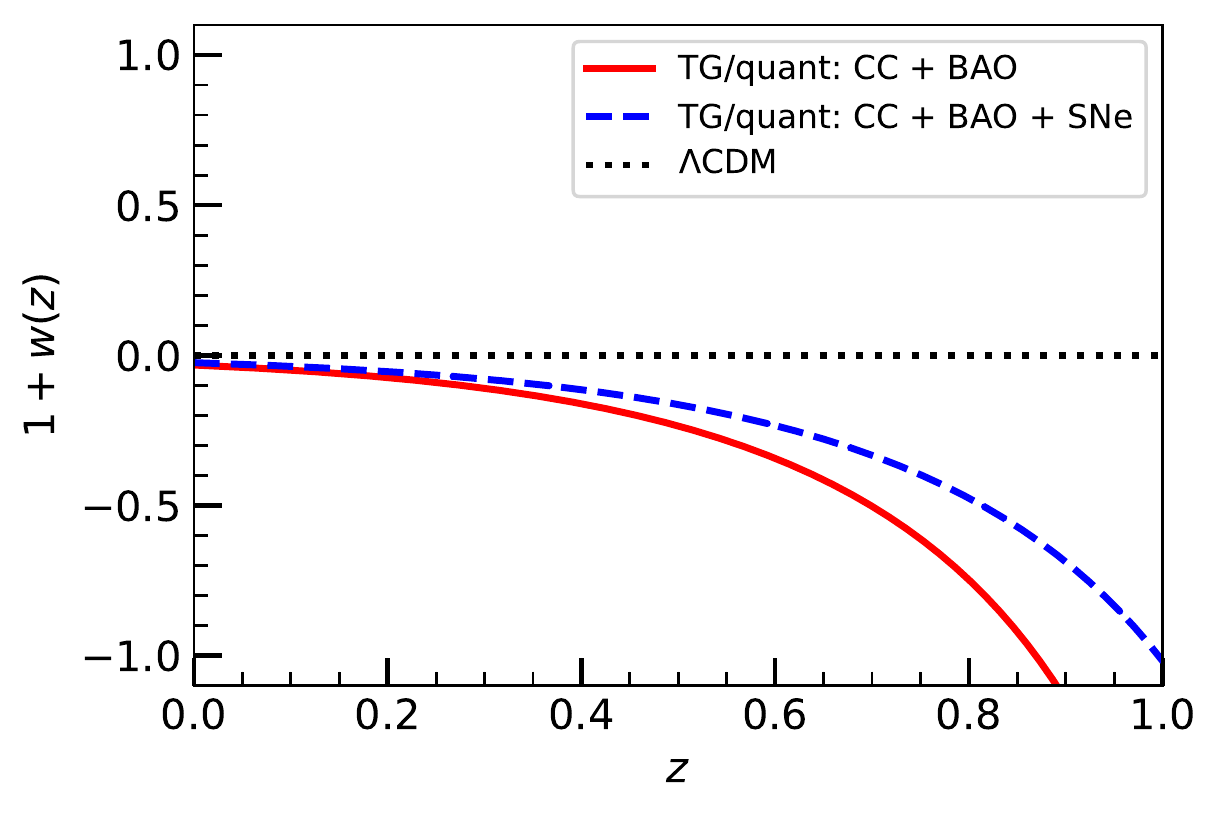}
\caption{Best fit curves of the effective equation of state $w$ as a function of redshift. The violation of null energy condition can be clearly identified.}
\label{fig:eos}
\end{figure}

Before closing this section, we would like to mention that the quantum corrections in this model in fact generate a slightly ``phantom-like" accelerating expansion of the Universe, even though there is no any exotic matter fields that violate null energy condition. This can be seen by defining an effective fluid whose energy density and pressure are respectively the collection of all terms, in the right-hand side of Eqs.~(\ref{eq:FeqM}) and (\ref{eq:PeqM}), subtracting the contributions from the baryonic and cold dark matter components. The best fit curve of the equation of state $w$ of this effective fluid, as a function of redshift, is shown in Figure \ref{fig:eos}, in which the violation of null energy condition ($w<-1$) can be clearly seen. This is consistent with the latest observational results of Planck 2018 \cite{Planck:2018vyg} that an equation of state smaller than $w = -1$ for the $w$CDM model is slightly preferred. Also, the violation of null energy condition in this scenario is purely quantum and no exotic matter is needed.

\section{Conclusion}
\label{sec:conclusions}

We studied a cosmological model which takes into account semi-classical quantum corrections in teleparallel gravity. In particular, we focused on its implications in the late Universe by using compiled data sets from CC, BAO, and SNe sources. The statistical results were in slight favor of the inclusion of quantum fluctuations with a density parameter of $\Omega_{q0} = -0.03 \pm 0.02$ (see Tables \ref{tab:bestfits} and \ref{tab:statistics}). This edge is not too significant to rule out the quantum fluctuations nor the bare $\Lambda$CDM model, but it merits the subject further study. This mild preference for quantum corrections may become more pronounced with future data sets.

The deviation from a $\Lambda$CDM background is only moderately changed in this quantum context. Despite this fact, the $\alpha$ parameter ($\sim 10^{32}$ seconds) takes on very large values unless standard cosmological units are used. This comes about due to the inheritance of units from the Hubble constant, as discussed after Eq.~(\ref{eq:Meq_pressureless}). In spite of this, the impact of these quantum corrections remains mild. We also stress that our physical estimate (a mean and an uncertainty) does not rule out small, or even Planck scale ($\sim 10^{-44}$ seconds), values of the quantum parameter. This is clear in Figure \ref{fig:post_oq0_q_TG} where these minute amounts are always within the ninety-five percent confidence contours. None of the statistical evidence is significant enough to claim a preference for large quantum parameters.

We clarify that the ``effective'' approach advocated in this work (and originally proposed in Refs. \cite{Dzhunushaliev:2013nea, Dzhunushaliev:2015mva, Chen:2021oal}) is based on a decomposition of the gravitational field, provided that its quantum part is much smaller than its classical part, or rather that $q \ll 1$, which is supported by our estimate based on cosmic expansion data. This led to non-minimal gravity-matter couplings which we tested with available data, as should be for any model. It is not necessary for $\alpha$ to be strictly Planck scale in this description, but understandably non-Planckian estimates of a quantum parameter should be subject to further inspection. Nonetheless, by and large, quantum phenomena manifesting in a classical setting are also not uncommon in physics, e.g., phantom-like dark energy sourced by quantum gravity \cite{Oriti:2021rvm} and stochastic noise influencing weak field systems \cite{Chawla:2021lop}, among others \cite{Parikh:2020nrd, Parikh:2020kfh, Parikh:2020fhy}. Of course, this is not intended to be a justification to overlook a faithful quantum gravity theory. Instead, while we search for it, we hope that such effective description together with observational data broaden our current understanding of the Universe.

It should be emphasized as well that a better fit to the background observation does not necessarily imply modified gravity \cite{Wen:2021bsc}. A better fit with data may just so happen to occur even if the underlying physics is governed by standard cosmology, in other words, unmodified gravity. This calls for a future work with perturbations, singling out smoking-gun observations sourced by quantum fluctuations. A competitive background fit is a good start.

A result also worth mentioning is that the inclusion of quantum corrections also catered to lower values of $H_0$. This may be useful to keep in mind in the background of the Hubble tension.

Various directions can be taken from here. There are a variety of other models presenting semi-classical corrections as modified gravity \cite{Dzhunushaliev:2013nea, Dzhunushaliev:2015mva, Chen:2021oal}. We have only tested so far a representative, novel model that does not submit to an alternative Riemannian description. The influence of these semi-classical, teleparallel gravity corrections on linear cosmological observables will also be interesting to see in a future work. On the other hand, the non-minimal matter couplings should affect geodesic motion, test masses and so could also potentially be constrained using gravitational waves. Lastly, it would be fascinating to lookout for these quantum effects in the highly nonlinear regime, e.g., black holes \cite{Yang:2020lxv}, and study whether they may leave observational imprints such as shadow modifications.

\begin{acknowledgments}\label{sec:acknowledgements}
This research has been carried out using computational facilities procured through the European Regional Development Fund, Project ERDF-080 ‘A Supercomputing Laboratory for the University of Malta’.
\end{acknowledgments}

\appendix
%
\section{An alternative derivation of the background equations} \label{sec:alt_derivation}

The perfect fluid equations (Eqs. (\ref{eq:FeqM}), (\ref{eq:PeqM}), and (\ref{eq:Meq})) can alternatively be obtained by considering a dynamical scalar field $\phi$ described by
\begin{equation}
\mathcal{L}_m = \dfrac{1}{2} \left( \partial_\mu \phi \right) \left( \partial^\mu \phi \right) - V\left(\phi\right) \,,
\end{equation}
where $V$ is an arbitrary potential, and then performing a change of variables from $\left( \phi, V \right)$ in the scalar field Lagrangian given in Ref.~\cite{Chen:2021oal} to $\left( \rho, P \right)$ through the relation
\begin{eqnarray}
\rho &=& \dot{\phi}^2/2 + V\left(\phi\right) \\
P &=& \dot{\phi}^2/2 - V\left(\phi\right) \,.
\end{eqnarray}
This can be considered to be simply an exact transformation, with two variables $\left( \phi, V(\phi) \right)$ to be traded for two variables $\left( \rho, P(\rho) \right)$, one which is sufficient for an analysis involving background cosmological observations. An alternative, independent route leading to the same equations is of course an affirmation of mathematical consistency.

However, we also warn that such mapping approaches demand extra care when dealing with perturbations. Faithful fluid actions, e.g., Schutz-Sorkin action, can be considered for reliability in such applications.


\providecommand{\href}[2]{#2}\begingroup\raggedright\endgroup

\end{document}